\documentclass[runningheads,a4paper]{llncs}

\usepackage{amsmath,amssymb}
\usepackage[usestackEOL]{stackengine}
\usepackage{graphicx}
\usepackage{caption}
\usepackage{subcaption}
\usepackage{color}
\usepackage{tikz}
\usepackage{url}
\usepackage{wrapfig}
\usepackage{multirow}
\usetikzlibrary{automata}
\usetikzlibrary[arrows,decorations.pathmorphing,backgrounds,positioning,fit,petri]
\usetikzlibrary{decorations.pathreplacing,intersections}
\usepackage{mathptmx}

\usepackage{algorithm}
\usepackage{algpseudocode}

\usepackage{array}
\newcolumntype{L}[1]{>{\raggedright\let\newline\\\arraybackslash\hspace{0pt}}m{#1}}
\newcolumntype{C}[1]{>{\centering\let\newline\\\arraybackslash\hspace{0pt}}m{#1}}
\newcolumntype{R}[1]{>{\raggedleft\let\newline\\\arraybackslash\hspace{0pt}}m{#1}}
\makeatletter
\def\BState{\State\hskip-\ALG@thistlm}
\makeatother

\tikzset{
  bend angle=30,
  -,
  shorten >=1pt,
  node distance=2cm and 2cm,
  on grid,
  auto,
  initial where=above,
  initial text=,
  initial distance=0.6cm,
  inner sep=0.5mm,
  smallstate/.style={circle,draw},
  openstate/.style={},
  succstate/.style={font={$\surd$}},
  bendanglelarge/.style={bend angle=45},
  bendanglesmall/.style={bend angle=10},
  dim/.style={lightgray},
  diag/.style={red},
  bisim/.style={red,dashed,-,bend left},
  loop left/.style={in=195,out=165,loop,swap}
}

\usepackage[colorlinks, linkcolor=black, anchorcolor=black,
	    citecolor=blue, urlcolor=black, bookmarks=true
            ]{hyperref}

\begin{document}

\mainmatter
\title{Parallel Approximate Steady-state Analysis of Large Probabilistic Boolean Networks (Technical Report)}

\titlerunning{Parallel Approximate Steady-state Analysis}


\author{Andrzej Mizera\inst{1} \and Jun Pang\inst{1,2} \and Qixia Yuan\inst{1}\thanks{Supported
by the National Research Fund, Luxembourg (grant 7814267). }}

\authorrunning{Mizera, Pang, and Yuan}


\institute{
University of Luxembourg, FSTC, Luxembourg
\and
University of Luxembourg, SnT, Luxembourg\\
\email{firstname.lastname@uni.lu}
}
\maketitle

\begin{abstract}
Probabilistic Boolean networks (PBNs) is a~widely used
computational framework for modelling biological systems.
The steady-state dynamics of PBNs is of special interest in the analysis
of biological systems.
However, obtaining the steady-state distributions for such systems poses a~significant challenge
due to the state space explosion problem which often arises in the case of large PBNs.
The only viable way is to use statistical methods.
We have considered the two-state Markov chain approach and the Skart method for the analysis of large PBNs in our previous work.
However, the sample size required in both methods is often huge in the case of large PBNs
and generating them is expensive in terms of computation time.
Parallelising the sample generation is an ideal way to solve this issue.
In this paper, we consider combining the German \& Rubin method with either the two-state Markov
chain approach or the Skart method for parallelisation. The first method can be used to run
multiple independent Markov chains in parallel and to control their convergence to the
steady-state while the other two methods can be used to determine the sample size required
for computing the steady-state probability of states of interest.
Experimental results show that our proposed combinations can reduce time cost
of computing stead-state probabilities of large PBNs significantly.


\end{abstract}

\section{Introduction}
\label{sec:intro}
Computational systems biology aims to model and analyse biological systems from a holistic
perspective with the use of formal, mathematical reasoning and computational techniques that
exploit efficient data structures and algorithms.
Computational modelling allows systematisation of available biological knowledge concerning
biochemical processes of a~biological system and provides formal means for the analysis and
understanding of real-life systems. Unfortunately, it often arises that the size of the state
space of the system to be considered is so huge that it prohibits the analysis.
Thus comprehensive understanding of biological processes requires further development of
efficient methods and techniques for formal modelling and analysis of biological systems.

One key aspect of analysing biological systems is to understand their long-run behaviour, which
is crucial in many contexts, e.g., in the analysis of the long-term influence of one gene on another
gene in a~gene regulatory network (GRN)~\cite{SDZ02}. In this work, we concentrate on the
steady-state analysis of biological mechanisms, in particular GRNs, cast into the framework of
probabilistic Boolean networks (PBNs). As introduced by Shmulevich et al.~\cite{SD10}
(see~\cite{TMPTSS13} for a~recent survey), PBNs are a~probabilistic generalisation of the standard
Boolean networks: they not only incorporate rule-based dependencies between genes and allow the
systematic study of global network dynamics; but also are able to deal well with uncertainty,
which comes naturally in biological processes. The dynamics of PBNs can be studied in the realm of
discrete-time Markov chains (DTMCs). Therefore, the rich theories of DTMCs can be applied to the
analysis of PBNs.

Given a~PBN, one natural and crucial issue is to study the steady-state probabilities of its
underlying DTMC, which characterise the long-run behaviour of the corresponding biological
systems~\cite{Kau93}. Much effort has been spent in analysing the steady-state behaviour of
biological systems for better understanding the influences of genes or molecules in the systems,
e.g., the ebb and flow of molecular events during cancer progression~\cite{SDZ02}. Furthermore, 
steady-state analysis has been used in gene intervention and external
control~\cite{SDZ02Gene,ADZ02,AD07,QD08},which is of special interest to cancer therapist to
predict the potential reaction of a~patient.

It has been well studied how to compute the steady-state probabilities of small-size PBNs using numerical methods~\cite{SGH03,PAJ14}.
However, in the case of large PBNs,
their state-space size becomes so huge that the numerical methods,
often relying on the transition matrix of the underlying DTMC of the studied network,
are not scalable any more.
This poses a critical challenge for the steady-state analysis of large PBNs.
In fact, approximations with Markov Chain Monte Carlo (MCMC) techniques remain the only feasible method to solve the problem.
In our previous work~\cite{MPY15}, we have considered the two-state Markov chain approach and the Skart method for
approximate analysis of large PBNs.
Taking special care of efficient simulation,
we have implemented these two methods in the tool \textsc{ASSA-PBN}~\cite{assa},
and successfully used it for the analysis of large PBNs with a few thousands of nodes.
However, the trajectory required for analysing a large PBN is often very long and
generating such long trajectories is expensive in terms of computation time.
A natural idea for speeding up this is to perform the trajectory generation in parallel.
In this work, we consider combining the Gelman \& Rubin method~\cite{GR92}
with the two methods we have considered in~\cite{MPY15}.
We simulate multiple trajectories in parallel and verify the convergence of the trajectories based
on the Gelman \& Rubin method. Once convergence is reached according to the Gelman \& Rubin
method, either the two-state Markov chain approach or the Skart method can be applied to the
converged trajectories to compute the steady-state probability for a~set of states that are of
interest. We show with experiments that the combinations can significantly reduce the computation
time for approximate steady-state analysis of large PBNs.


\section{Preliminaries}
\label{sec:preliminaries}

\subsection{Finite discrete-time Markov chains}
\label{ssec:dtmc}
We define a discrete-time Markov chain (DTMC) as a tuple ($S, s_0, P$),
where  $S$ is a finite set of states,  $s_0 \in S$ is the initial state, and $P:  S \times S \rightarrow [0,1]$ is a transition probability matrix.
For any two states $s$ and $s'$, an element of $P(s,s')$ defines the probability
that a transition is made from state $s$ to state $s'$.
It satisfies that $P(s,s') \geq 0$ for all $s,s'\in S$ and $\sum_{s' \in S} P(s,s')=1$ for all
$s\in S$. A path with length $n$ is a sequence of states $s_0, s_1, \ldots, s_{n-1}$,
where $s_i \in S$ for all $i\in \{0,1,\ldots,n-1\}$ and $P(s_i,s_{i+1})>0$ for all $i\in \{0,1,\ldots,n-2\}$.
State $s'\in S$ is said to be \emph{reachable} from state $s\in S$ if there exists a~path from $s$ to $s'$.
A DTMC is said to be \emph{irreducible} if any two states in the state space are reachable from each other.
A state of a DTMC is of period $d$ such that $d$ equals to the greatest common divisor of the lengths of all paths that start and end
in the state.
If all states in the state space of a DTMC are of period one, then the DTMC is \emph{aperiodic}.
A~finite state DTMC that is both irreducible and aperiodic is \emph{ergodic}.
Let $\pi$ be a~probability distribution on $S$.
$\pi$ is called a \emph{stationary distribution} of the DTMC if $\pi = \pi\cdot P$.
According to the ergodic theory of DTMC~\cite{NORRIS98},
an ergodic DTMC has a unique stationary distribution,
being simultaneously its \emph{limiting distribution}.
It is also known as the \emph{steady-state distribution}
given by $\lim_{n\to\infty}\pi_0\cdot P^n$, where $\pi_0$ is any initial probability distribution on $S$ and $P^n$ is the $n$ times multiplication of the transition matrix $P$.
Therefore, the limiting distribution of an ergodic DTMC is independent of the choice of the initial distribution
and it can be estimated by iteratively multiplying $P$.

\subsection{Probabilistic Boolean network}
\label{ssec:pbn}
A~probabilistic Boolean network $G(V,\mathcal{F})$ is composed of a~set of binary-valued variables
(also referred to as nodes) $V=(v_{1},v_{2}, \ldots ,v_{n})$ whose values are governed by a list
of sets $\mathcal{F}=(F_{1},F_{2}, \ldots, F_{n})$.
The set $F_{i}=\{f_{1}^{i},f_{2}^{i},\ldots,f_{\ell(i)}^{i}\}$ is defined as a set of possible predictor
functions for node $v_{i}$, where $i \in \{1,2,\ldots,n\}$ and $\ell(i)$ is the number of possible
predictor functions for $v_{i}$.
Each predictor function $f_{j}^{i}$ is a~Boolean function defined with respect to a~subset of
nodes referred to as parent nodes of the node $v_i$.
At a given time point $t$ ($t=0,1,\ldots$), one predictor function is selected for each of the nodes.
We call the combination of all the selected prediction functions at time $t$ a \emph{realisation} of a PBN.
Assuming independence among the predictor functions for different nodes,
there are $N=\prod_{i=1}^{n}\ell(i)$ possible realisations for a PBN.
We denote the realisations as vectors $\boldsymbol{f}_k$, $k\in \{1,\ldots,N\}$,
where the $i$-th element is the Boolean function selected for node $v_i$.
The realisation at time point $t$ is expressed as $f(t)$.
For a node $v_i$, the selection probability for selecting its $j$th predictor function is denoted as
$c_j^{(i)}$ and it holds that  $\sum_{j=1}^{\ell(i)}c_{j}^{(i)}=1$.
The state of a PBN at time point $t$, denoted as $s(t)$, is defined as a collection of all the
node values at time $t$,
namely $s(t)=(v_1(t),v_2(t),\ldots,v_n(t))$, where $v_i(t)$ is the value of node $v_i$ at time $t$.
A~PBN with $n$ nodes has $2^n$ possible states and $\boldsymbol{s}(t) \in \{0,1\}^{n}$ for each
$t$.
The transition from $s(t)$ to $s(t+1)$ is conducted by synchronously updating the node values according to the realisation at time $t$,
i.e., $s(t+1)=f(t)(s(t))$.

The concept of perturbations is introduced to PBN by providing a parameter
$p \in (0,1)$, which is used to sample a perturbation vector $\boldsymbol{\gamma}(t)=(\gamma_{1}(t),\gamma_{2}(t),\ldots,\gamma_{n}(t))$,
where each $\gamma_i(t)\in \{0,1\}$ is a~Bernoulli distributed random variable with parameter $p$
for all $t$ and $i\in \{1,2, \ldots,n\}$. If $\gamma_i(t)=0$,
the next state of a PBN is given by $s(t+1)=f(t)(s(t))$;
otherwise, it is determined as $\boldsymbol{s}(t+1)=\boldsymbol{s}(t) \oplus \boldsymbol{\gamma}(t)$,
where $\oplus$ is the `exclusive or' operator for vectors.
Perturbations allow to reach an arbitrary state from any other state within one transition in a PBN.
Thus the dynamics of a PBN with perturbations can be viewed as an ergodic DTMC over $S=\{0,1\}^n$~\cite{SD10}.
With the ergodic theory of DTMCs~\cite{NORRIS98}, it can be concluded that the long run dynamics
of a~PBN is independent of the choice of its initial state.
This allows the estimation of the steady-state behaviour of a PBN by performing simulation from
an~arbitrary initial state.

The density of a~PBN is measured with its predictor functions number and parent nodes number.
For a~PBN $G$, its density is defined as $\mathcal{D}(G)=\frac{1}{n}\sum_{i=1}^{\it 
N_{F}}\theta(i)$, where ${\it N_{F}}$ is the total number of predictor functions in $G$ and
$\theta(i)$ is the number of parent nodes of the $i$-th predictor function.

\section{Steady-state Analysis of PBNs}
\label{sec:ssa}
As shown in~\cite{MPY15}, both the two-state Markov chain approach and the Skart method
are effective for analysing steady-state probabilities of a~PBN with number of nodes up to a~few
thousands. We briefly discuss in this section these two methods.
An efficient implementation of the two methods for the
analysis of large PBNs is available in the ASSA-PBN tool~\cite{assa}.

\subsection{The two-state Markov chain approach}
\label{ssec:two}
\begin{figure}[!t]
  \centering
  \begin{subfigure}[b]{0.48\textwidth}
    \centering
    \includegraphics[scale=0.35]{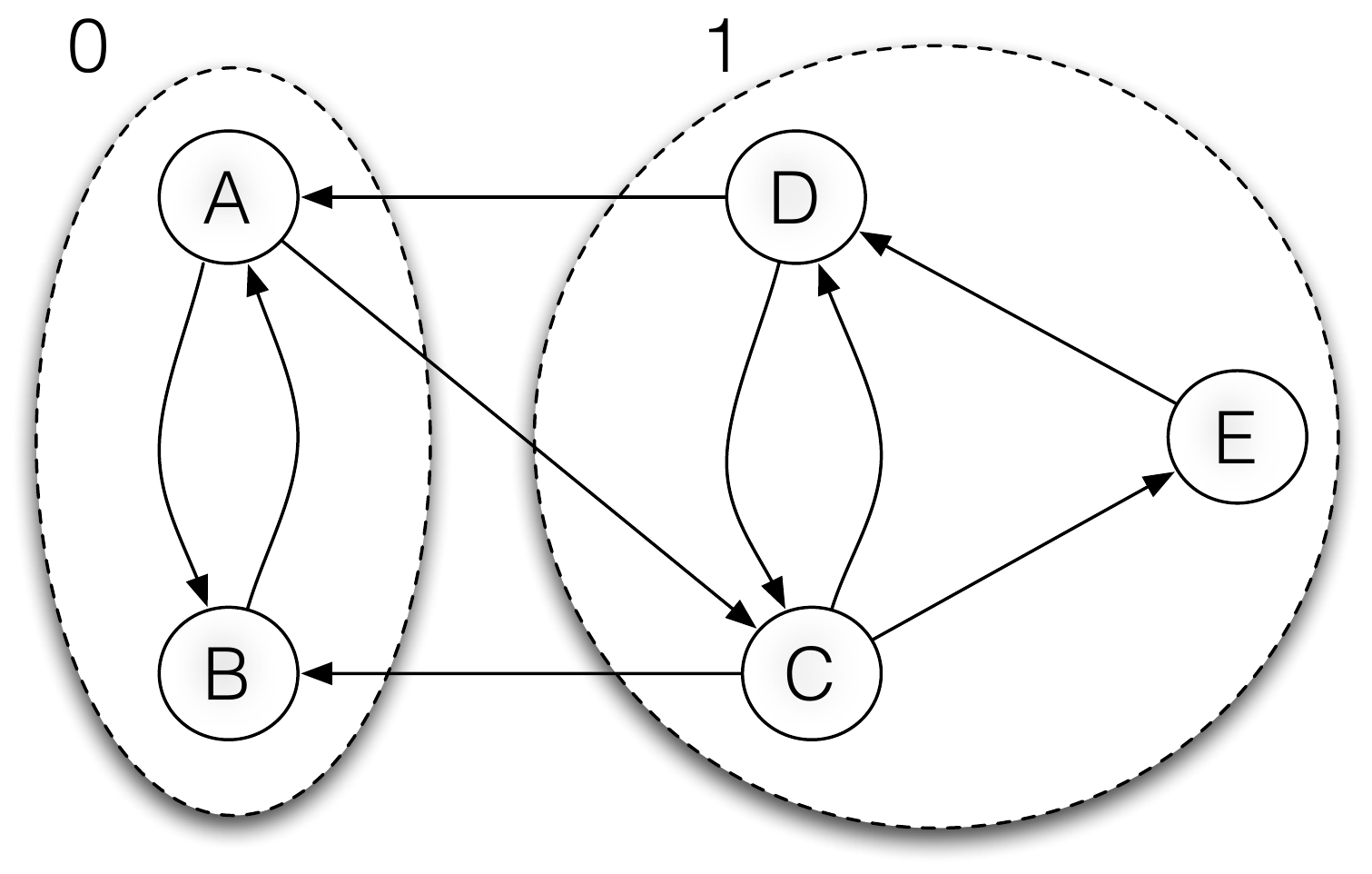}
    \caption{Original DTMC}
    \label{fig:dtmc}
  \end{subfigure}%
  \quad
  \begin{subfigure}[b]{0.48\textwidth}
    \centering
    \includegraphics[scale=0.35]{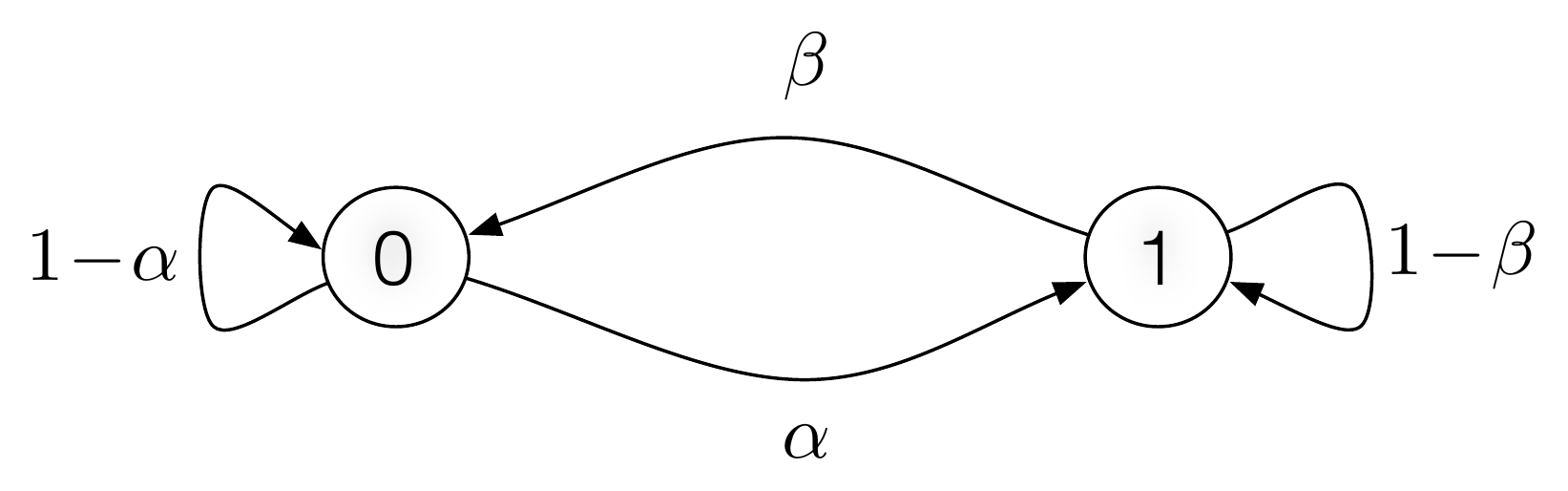}
    \vspace{5mm}
    \caption{Two-state DTMC }
    \label{fig:two-state_MC}
  \end{subfigure}
  \caption{Conceptual illustration of the idea of the two-state Markov chain construction. (a) The
  state space of the original discrete-time Markov chain is split into two meta states: states
  $A$ and $B$ form meta state~$0$, while states $D$, $C$, and $E$ form meta state~$1$. The split
  of the state space into meta states is marked with dashed ellipses. (b) Projecting the behaviour
  of the original chain on the two meta states results in a~binary (0-1) stochastic process which
  can be approximated as a~first-order, two-state Markov chain.}
  \label{fig:two-state_MC_approach}
\end{figure}

The two-state Markov chain approach~\cite{RL92} is a~method for approximate computation of the
steady-state probability for a~subset of states of a~DTMC.
This approach splits the states of an arbitrary DTMC into two parts, referred to as two meta
states. One part is composed of the states of interest, numbered $1$, and the other part is its
complement, numbered $0$.
Such consideration abstracts an arbitrary DTMC into a~0-1 stochastic process, which can further be
approximated by a~first-order, two-state DTMC that consists of the two meta states with transition
probabilities $\alpha$ and $\beta$ between them.
Fig.~\ref{fig:two-state_MC_approach} illustrates the construction of a two-state Markov chain from
a~5-state Markov chain.

The steady-state probability of meta state $1$ can be estimated by performing simulation of the original DTMC.
This estimation is achieved iteratively using the standard two-state Markov chain approach of~\cite{RL92} to guarantee
that the samples used for estimation are drawn from a distribution which differs from the
the steady-state distribution at most by $\epsilon$ and that two precision requirements (precision
$r$, and confidence level $s$) are satisfied.
We outline the steps in Algorithm~\ref{alg:two-state_MC}.
The two arguments $m_0$ and $n_0$ are the initial `burn-in'
period and the initial sample size, respectively.
In each iteration of the algorithm,
the `burn-in' steps $M$ and the actual sample size $N$ are re-estimated.
The iteration continues until the estimated sample size ($M+N$) is not bigger than the current trajectory length.
For more details on this approach,
a~derivation of the formulas for $M$ and $N$ in line~\ref{lst:M_N}, and a~discussion regarding
a~proper choice of $n_0$, we refer to~\cite{MPY15}.

\begin{algorithm}[!t]
\caption{The Two-state Markov chain approach}
\label{alg:two-state_MC}
\begin{algorithmic}[1]
\Procedure{estimateProbability}{$m_0,n_0,\epsilon,r,s$}
    \State $M := m_0$; $N := n_0$; $l := M + N$;
    \State Generate an~initial trajectory of length $l$ abstracted to the two meta states.
    \Repeat
    \State Extend the trajectory by $M+N-l$.
    \State $l := M + N$;
    \State Estimate $\alpha,\beta$ based on the last $N$ elements of the extended trajectory
    \State  $M := \left\lceil \log{\left(\frac{\epsilon(\alpha+\beta)}{\max(\alpha,\beta)}\right)}/
     \log{(|1-\alpha-\beta|)}\right\rceil
     N := \left\lceil \frac{\alpha\beta(2-\alpha-\beta)}{(\alpha+\beta)^3}
     \frac{\left(\mathrm{\Phi}^{-1}(\frac{1}{2}(1+s))\right)^2}{r^2}\right\rceil$
     \label{lst:M_N}
	  \Until{$M+N\leq l$}
	\State Estimate the prob. of meta state 1 from the last $N$ elements of the trajectory.
\EndProcedure
\end{algorithmic}
\end{algorithm}

\subsection{The Skart method}
\label{ssec:skart}
\begin{figure}[!t]
  \centering
  \begin{subfigure}[b]{0.48\textwidth}

    \centering
\begin{tikzpicture}
   \draw[thick] (0,0) --  (4.5,0);    
   \foreach \y in {0,1,2,3.5,4.5}{
     \draw (\y,0) -- (\y,.2);
     }                              
   \draw [decorate,decoration={brace,raise=1pt}]
      (0,0.2) -- (1,0.2) node [midway,yshift=5] {$\kappa$};    
    \draw [decorate,decoration={brace,raise=1pt}]
      (1,0.2) -- (2,0.2) node [midway,yshift=5] {$\kappa$};    
      \draw [decorate,decoration={brace,raise=1pt}]
      (3.5,0.2) -- (4.5,0.2) node [midway,yshift=5] {$\kappa$};    
      \draw [decorate,decoration={brace,raise=1pt}]
     (4.5,-0.1)--  (0,-0.1)  node [midway,yshift=-5] {$\eta=\kappa*p$};    
      \node[align=right] at (2.7,0.3) {...};
  \end{tikzpicture}
  \caption{A single chain is divided into $p$ batches}
   \label{fig:singlechain}
    \end{subfigure}%
  \quad
  \begin{subfigure}[b]{0.48\textwidth}

    \centering
    \begin{tikzpicture}
   \draw[thick] (0,0) --  (4.5,0);    
   \foreach \y in {0,1,2,3.5,4.5}{
     \draw (\y,0) -- (\y,.2);
     }                              
   \draw [decorate,decoration={brace,raise=1pt}]
      (0,0.2) -- (1,0.2) node [midway,yshift=5] {$\kappa$};    
    \draw [decorate,decoration={brace,raise=1pt}]
      (1,0.2) -- (2,0.2) node [midway,yshift=5] {$\kappa$};    
      \draw [decorate,decoration={brace,raise=1pt}]
      (3.5,0.2) -- (4.5,0.2) node [midway,yshift=5] {$\kappa$};    
      \draw [decorate,decoration={brace,raise=1pt}]
     (4.5,-0.1)--  (0,-0.1)  node [midway,yshift=-5] {$\kappa*\lceil p/\omega\rceil$};    
      \node[align=right] at (2.7,0.3) {...};
      \node[align=right] at (0.3,1.1) {...};
       \draw[thick] (0,2) --  (4.5,2);    
   \foreach \y in {0,1,2,3.5,4.5}{
     \draw (\y,2) -- (\y,2.2);
     }                              
   \draw [decorate,decoration={brace,raise=1pt}]
      (0,2.2) -- (1,2.2) node [midway,yshift=5] {$\kappa$};    
    \draw [decorate,decoration={brace,raise=1pt}]
      (1,2.2) -- (2,2.2) node [midway,yshift=5] {$\kappa$};    
      \draw [decorate,decoration={brace,raise=1pt}]
      (3.5,2.2) -- (4.5,2.2) node [midway,yshift=5] {$\kappa$};    
      \draw [decorate,decoration={brace,raise=1pt}]
     (4.5,1.9)--  (0,1.9)  node [midway,yshift=-5] {$\kappa*\lceil p/\omega\rceil$};    
      \node[align=right] at (2.7,2.3) {...};

      \draw [decorate,decoration={brace,raise=1pt}]
      (0,0) -- (0,2.2) node [midway,xshift=-5] {$\omega$};    
  \end{tikzpicture}
  \caption{$\omega$ chains are divided into $p$ batches}
    \label{fig:multiplechain}
  \end{subfigure}
  \caption{Demonstration of dividing samples into batches for the Skart method.
  (a) Dividing samples from a single chain into $p$ batches, each of size $\kappa$.
  The chain size $\eta=\kappa*p$.
  (b) Dividing samples from $\omega$ chains into $p$ batches, each of size $\kappa$.
  The size of each chain is $\kappa* \lceil p/\omega\rceil$.
  The actual number of batches after dividing is $p'=\omega*\lceil p/\omega\rceil$.}
  \label{fig:divide}
  \end{figure}
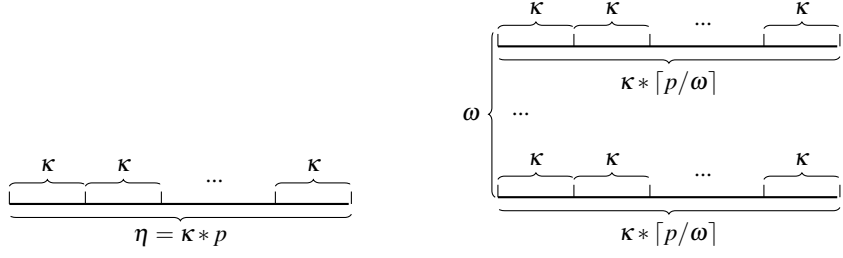

Proposed by Tafazzoli et al.~\cite{AJE08} in 2008,
the Skart method is a non-overlapping batch mean method
that can be used to estimate the steady-state probabilities of a DTMC
from a simulated trajectory of the DTMC.
It divides the simulated trajectory of length $\eta$, i.e., $\{X_i:i=1,2,\ldots,\eta\}$,  into $p$ non-overlapping batches, each of size $\kappa$.
See Figure~\ref{fig:singlechain} for an~illustration.
Assuming $p$ and $\kappa$ are both large enough,
it guarantees that the batch means are approximately independent and identically distributed (i.i.d) normal random variables.
The grand mean of the individual batch means, denoted as $\mu$,
is considered as a point estimator of  $\mu_{X}$, i.e., the expected value of $X_i$.
In practise, some initial batches, known as `burn-in' steps and denoted as $\zeta$, are discarded to eliminate the initialisation bias when computing the point estimator $\mu_{X}$.
The method then constructs a CI (confidence interval) estimator  for $\mu_{X}$ that is centered on $\mu$.
The key process of the Skart method is to determine a proper batch size $\kappa$, a proper batch number $p$, and proper number of `burn-in' steps $\zeta$,
so that the computed steady-state estimations are approximately the theoretical ones and the computed CI estimator satisfies certain precision requirements.
This is achieved by using randomness test, autocorrelation, and skewness adjustment in an iterative way.
We summarise the process of the Skart method in Algorithm~\ref{alg:skart}, and we refer to~\cite{AJE08} for a more detailed description.
Given two input parameters $H^*$ (precision requirement) and  $\alpha$ (confidence level),
this algorithm computes a CI estimator $[CI_{bottom}, CI_{top}]$ and a point estimator $\mu$ which together satisfy that $H^*<max\{\mu-CI_{bottom},CI_{top}-\mu\}$
and the real steady-state probability of the system is within $100(1-\alpha)\%$ probability.

\begin{algorithm}[!t]
\caption{The Skart method}
\label{alg:skart}
\begin{algorithmic}[1]
\Procedure{estimateCI}{$H^*, \alpha$}
    \State $\eta:=1280$; $l:=1024$; $pass:=\textsc{false}$;
    \State Generate an~initial trajectory of length $\eta$.
    \State Compute the skewness $\hat{B}$ of the last $l$ samples and set batch size $\kappa$ based on $\hat{B}$.
   \State $p:=\eta$; $\eta:=\kappa*p$;
   \State Extend trajectory to $\eta$ and compute randomness test statistics $C$
   \While{Independence test is not passed}
    \State Adjust batch size $\kappa$, number of batches $p$ and spacer~\cite{FGW91} size $d$; $\eta:=\kappa*p$
     \State Extend trajectory to $\eta$ and compute randomness test statistics $C$
  \EndWhile
	\State $\zeta:=d*\kappa$; skip first $\zeta$ samples.
	\Repeat
	\State Extend the trajectory to length $\eta$ if necessary.
	\State Compute nonspaced batch means $\mu$ and variance estimator $Var$.
	\State Compute skewness and autogression adjusted $CI$ based on $Var$ and $\alpha$.
    \State $H=max\{\mu- CI_{bottom},CI_{top}-\mu\}$;
    \If{$H > H^*$} \hfill{\it //check whether the precision requirement is satisfied}
    \State Adjust batch size $\kappa$ and number of batches $p'$; \State  $\eta:=\kappa*(p'+d)$
     \hfill{\it //$p'$ does not contain the number of discarded batches}
    \Else~ $pass:=\textsc{true}$;
    \EndIf
	\Until{$pass$}
\EndProcedure
\end{algorithmic}
\end{algorithm}

\section{Parallel Steady-state Analysis of Large PBNs}
\label{sec:parallel}
As shown in~\cite{MPY15}, the two above mentioned  simulation-based methods work well for
computing the steady-state probabilities for large PBNs even with a few thousands of nodes.
However, it often appears that a huge sample size is required in the case of large PBNs,
which can be computationally expensive.
In principle, parallelising the sample generation process can be considered an~ideal solution to
this problem.
We propose to combine the Gelman \& Rubin method with the two above mentioned methods.
The Gelman \& Rubin method is used to monitor that all the simulated chains have approximately
converged to the steady-state distribution while the other two methods are used to determine the
sample size required for computing the steady-state probabilities of the states of interest.
In the following part of this section,
we first introduce the Gelman \& Rubin method and then discuss how to parallelise the two methods
mentioned in Section~\ref{sec:ssa} by using the Gelman \& Rubin method.

\subsection{The Gelman \& Rubin method}
\label{ssec:gr}
The Gelman \& Rubin method~\cite{GR92} is an~approach for monitoring the convergence of multiple
chains.
It starts from simulating $2\psi$ steps of $\omega \geq 2$ independent Markov chains in parallel.
The first $\psi$ steps of each chain, known as the `burn-in' period, are discarded from it.
The last $\psi$ elements of each chain are used to compute the within-chain ($W$) and between-chain ($B$) variance,
which are used to estimate the variance of the steady state distribution ($\hat{\sigma}^2$).
Next, the potential scale reduction factor $\hat{R}$ is computed with $\hat{\sigma}^2$.
$\hat{R}$ indicates the convergence to the steady state distribution.
The chains are considered as converged and the algorithm stops if $\hat{R}$ is close to 1;
otherwise, $\psi$ is doubled, the trajectories are extended, and $\hat{R}$ is recomputed.
We list the steps of this approach in Algorithm~\ref{alg:GR}. For further details of this method and the
discussion on the choice of the initial states for the $\omega$ chains we refer to~\cite{GR92}.

\begin{algorithm}[!t]
\caption{The Gelman \& Rubin method}
\label{alg:GR}
\begin{algorithmic}[1]
\Procedure{generateConvergedChains}{$\omega,\psi_0$}
  \State $\psi := \psi_0$;
  \State Generate in parallel $\omega$ trajectories of length $2\psi$;
  \Repeat
    \State chains(1\ldots$\omega$,1\ldots$2\psi$) := Extend all the $\omega$ trajectories to length $2\psi$;
    \For{$i=1..\omega$}
      \State $\mu_i := $ mean of the last $\psi$ values of chain $i$;
      \State $s_i := $ standard deviation of the last $\psi$ values of chain $i$;
    \EndFor
    \State $\mu := \frac{1}{\omega}\sum_{i=1}^\omega \mu_i$;
    \State $B:=\frac{\psi}{\omega-1}\sum_{i=1}^{\omega}(\mu_i-\mu)^2;
      \ W:=\frac{1}{\omega}\sum_{i=1}^{\omega}s_{i}^{2}$;
      \hfill{\it //Between and within variance}
    \State $\hat{\sigma}^2 :=(1-\frac{1}{\psi})W+\frac{1}{\psi}B$;
      \hfill{\it //Estimate the variance of the stationary distribution}
    \State $\hat{R} := \sqrt{\hat{\sigma}^2/W}$;
      \hfill{\it //Compute the potential scale reduction factor}
    \State $\psi := 2\cdot\psi$;
  \Until{$\hat{R}$ is close to 1}
  \State \Return (chains,$\psi/2$);
\EndProcedure
\end{algorithmic}
\end{algorithm}

\subsection{Parallelising the two-state Markov chain approach}
\label{ssec:parallelTwostate}
To reduce the time cost of the two-state Markov chain approach in the case of large PBNs,
we propose to parallel this approach by providing samples from multiple chains.
This is achieved by combing the two-state Markov chain approach with the German \& Rubin method.
The German \& Rubin method is used to run multiple chains of the original DTMC
to assure their convergence to the steady-state distribution.
Once convergence is reached, the second halves of the chains are merged into one sample,
and the two-state Markov chain approach is applied to estimate $N$ based on
the merged sample.
As convergence is guaranteed, no `burn-in' period is considered.
The stop criterium for the two-state Markov chain approach becomes
that the estimated value $N$ is not bigger than the size of the merged sample.
If the stop criterium is not satisfied, the multiple chains are extended in parallel to provide a sample of required length.

The above idea is based on the assumption that once the simulated chains of the original Markov
chain have converged, the two-state Markov chain abstraction is also converged to its steady-state
distribution.
To guarantee that this assumption is correct, we add an~additional step.
Once the stop criterium is satisfied, the `burn-in' period $M$ of
the two-state Markov chain is computed.
The assumption is verified true if $M$ is not larger than the `burn-in' period $\psi$ of
the Gelman \& Rubin method.
Otherwise, additional $M-\psi$
elements will be discarded in the beginning of each chain and the two-state Markov chain part
is re-run on the modified sample. The detailed steps of this approach are outlined in
Algorithm~\ref{alg:combined}.

\begin{algorithm}[!t]
\caption{The Parallelised two-state Markov chain approach}
\label{alg:combined}
\begin{algorithmic}[1]
\Procedure{estimateProbabilityInParallel}{$\omega,\psi_0,\epsilon,r,s$}
  \State $(chains,\psi) := {\it generateConvergedChains}(\omega,\psi_0)$;
  \State $n := 0$; $extend\_by := \psi$;  $monitor :=\textsc{false}$; $abstracted\_sample :=\textsc{null}$;
  \Repeat
    \Repeat
      \State $chains := $ Extend in parallel each chain in chains by $extend\_by$;
      \State $sample := chains(1\ldots\omega,(n+\psi+1)\ldots(n+\psi+extended\_by))$;
      \State $abstracted\_sample := $ abstract $sample$ and combine with $abstracted\_sample$ ;
      \State $n :=n+ extend\_by$; $sample\_size := \omega\cdot n$;
      \State Estimate $\alpha,\beta$ from $abstracted\_sample$;
      \State Compute $N$ as Line 8 Algorithm~\ref{alg:two-state_MC};
      \State $extend\_by := \lceil (sample\_size - N)/\omega \rceil$;
    \Until{$extend\_by < 0$}

     \State Compute $M$ as Line 8 Algorithm~\ref{alg:two-state_MC}; $monitor := \textsc{false}$;
    \If{$M \geq \psi$}
      \State $extend\_by := \psi-M$; $\psi := M$; $monitor := \textsc{true}$;
    \EndIf
  \Until{$monitor$}
	\State Estimate the prob. of meta state 1 from $abstracted\_sample$;
\EndProcedure
\end{algorithmic}
\end{algorithm}

When analysing a~biological system, we are often interested in more than one set of states,
e.g., in the case of a~long-run sensitivity analysis of a~PBN used to model a~biological system.
For simplicity, we call the steady-state probability of the states of interest as one property
and computing this steady-state probability as checking one property.
Given $q$ different properties,
the two-state Markov chain approach needs to be run for $q$ times in order to check all of them.
Since the generation process is the most time consuming part in the algorithm,
the time cost for checking multiple properties can be reduced significantly if we can reuse the generated samples.
We modify Algorithm~\ref{alg:combined} to allow the reuse of samples for computing the steady-state probabilities
of multiple properties.
The crucial idea is that the simulated samples are abstracted into different meta states based on
these different $q$ properties simultaneously each time when an extension of chains is finished.
The calculations of $N$ for different properties are then performed simultaneously as well,
resulting in another level of parallelisation.
The next extension length is determined by the minimal value of all the calculated $N$s.
Using the minimal value would increase the extension times;
however it can avoid unnecessary abstraction of samples, which is a relatively expensive process.
The extension process is stopped when all the calculated $N$s are smaller than the current sample size.
The steady-state probabilities of all the properties are then calculated based on their
corresponding abstracted meta states.

\subsection{Parallelising the Skart method}
\label{ssec:parallelSkart}
The trajectory required by the Skart method can be very long as well in the case of large networks.
We propose to apply a similar strategy as what we have done for the two-state Markov chain approach to reduce the time cost of the Skart method.
It is assumed in the Skart method that the number of batches $p$ and the batch size $\kappa$
are large enough to guarantee that the batch means are approximately i.i.d normal random variables.
This assumption still holds when the batches are obtained from different chains given that convergence has been reached in those chains.
Therefore, the Skart method can be parallelised by fetching samples from multiple chains which have converged.
We use the Gelman \& Rubin method to guarantee that different chains have converged and
the Skart method to determine the number of batches and the size of each batch in order to estimate the target stationary probability with a given precision.
Since the burn-in steps are already discarded by the Gelman \& Rubin method,
no samples will be truncated when computing the CI in the parallelised version of the Skart method.
For efficiency consideration,
we further require that the number of batches obtained from each chain is the same.
This leads to the fact that the number of batches used in the parallelised Skart method is slightly larger than
that in the original Skart method. Figure~\ref{fig:divide} shows how a single chain and multiple chains are divided into batches.
We summarise the parallelisation of the Skart method in Algorithm~\ref{alg:skartparallel}
and highlight the lines where there exists a main difference with respect to the sequential method
by adding comments.

\begin{algorithm}[!t]
\caption{The Parallelised Skart method}
\label{alg:skartparallel}
\begin{algorithmic}[1]
\Procedure{estimateCIInParallel}{$H^*, \alpha$}
	\State $(chains,\psi):=generateConvergedChains(\omega,\psi_0);$
	\State Skip first $\psi$ samples of each chain; $\eta:=\lceil 1,280/\omega \rceil*\omega$; $pass:=\textsc{false}$;
    \State Extend all the chains each to length $\eta/\omega$ if necessary
    \State Compute the skewness $\hat{B}$ and set batch size $\kappa$ based on $\hat{B}$ \hfill{\it// use all samples}
    \State $p:=\lceil \eta/\omega\rceil*\omega $; $\eta:=\kappa*p$; \hfill{\it// batch number is adjusted}
   \State Extend all the chains each to $\eta/\omega$ and compute randomness test statistics $C$
   \While{Independence test is not passed}
    \State Adjust batch size $\kappa$, number of batches $p$ and spacer size $d$;
    \State $p:=\lceil p/\omega\rceil *\omega$; $\eta:=\kappa*p$\hfill{\it// batch number is adjusted}
     \State Extend all the chains each to $\eta/\omega$ and compute randomness test statistics $C$
  \EndWhile
	\Repeat
	\State Extend all the chains each to length $\eta/\omega$ if necessary
	\State Compute nonspaced grand batch mean $\mu$ and variance estimator $Var$
	\State Compute skewness and autogression adjusted $CI$ based on $Var$ and $\alpha$
    \State $H=max\{\mu- CI_{bottom},CI_{top}-\mu\}$
    \If{$H > H^*$}
    \State Adjust batch size $\kappa$ and number of batches $p$;
    \State  $p:=\lceil p/\omega\rceil *\omega$; $\eta:=\kappa*p$
     \hfill{\it // do not consider discarding the first $d$ samples}
    \Else~ $pass:= \textsc{true}$
    \EndIf
	\Until{$pass$}
\EndProcedure
\end{algorithmic}
\end{algorithm}

\section{Evaluation}
\label{sec:results}
We have implemented Algorithm~\ref{alg:two-state_MC} and Algorithm~\ref{alg:skart} in the tool \textsf{ASSA-PBN}~\cite{assa}
and evaluated their performance in our previous work~\cite{MPY15}.
We show in this section that the proposed two parallel algorithms
can significantly reduce the time cost for computing steady-state probabilities of large PBNs in comparison with their sequential versions.
We evaluate this first on randomly generated PBNs and then on a PBN from a real biological system.
All the experiments in this paper are conducted in a high-performance computing (HPC) node,
which contains 16 Intel Xeon E7-4850 processors@2GHz.
The 16 processors are equally distributed in 4 Bull S6030 boards (servers)
and each processor contains 10 cores.
This hardware architecture allows us to run a program with maximum 40 cores in one board.

\textsf{ASSA-PBN} was implemented in Java and the initial and the maximum Java Heap Size were set
to 2GB and 64GB, respectively. All the experimental data are available at
\url{http://satoss.uni.lu/software/ASSA-PBN/}.

\subsection{Speed-up for checking a single property}
\label{ssec:speeduponeproperty}
We first evaluate the speed-up for checking a single property using the parallelised algorithms, i.e., Algorithm~\ref{alg:combined} and Algorithm~\ref{alg:skartparallel}.
We randomly generate 18 different PBNs using the tool \textsf{ASSA-PBN}.
\textsf{ASSA-PBN} can randomly generate a PBN which satisfies
structure requirements given in the form of five input parameters: the node number, the
minimum and the maximum number of predictor functions per node, finally the minimum
and the maximum number of parent nodes for a predictor function.
The 18 PBNs are generated with node numbers from the set $\{80, 100, 200, 500, 1000, 2000\}$.
For each node number, 3 PBNs are generated.
We assign the obtained PBNs into three different classes with respect to their density
measure $\mathcal{D}$: dense models with density $150-300$, sparse models with density around
10, and in-between models with density $50-100$.
The precision and confidence level of all the experiments are set to $10^{-4}$ and 0.95, respectively.
The parameter $\epsilon$ in the two-state Markov chain approach is set to $10^{-10}$.
We compute steady-state probabilities for the 18 PBNs using
Algorithms~\ref{alg:two-state_MC},~\ref{alg:skart},~\ref{alg:combined}, and~\ref{alg:skartparallel}
and compare their results as well as time costs.
The parallelised algorithms are launched with 6 different number of cores in one board ranging
in $\{2, 5, 10, 20, 30, 40\}$.

As the models we use are too large to be dealt with numerical methods,
it is not possible to check the correctness of the parallelised algorithms using the models' theoretical steady-state probability distributions.
Instead we compare the results of the parallelised algorithms with their corresponding sequential algorithm.
We collect the two probabilities computed by Algorithms~\ref{alg:two-state_MC} and~\ref{alg:combined}
or by Algorithms~\ref{alg:skart} and~\ref{alg:skartparallel}
for checking the same property of one model as a pair.
In this section, there are 216 pairs of probabilities in total.
We expect the difference of the two probabilities in a pair to be less than $2 \times 10^{-4}$
with the probability of $95\%$.
In all the 216 pairs that we have obtained, the difference is always less than $2 \times 10^{-4}$.
Moreover, we perform  a similar verification for the evaluation results in Section~\ref{ssec:speedupmultiproperties}
and obtain the same observations.

\begin{figure}[!t]
  \centering
  \begin{subfigure}[b]{0.48\textwidth}
    \centering
    \includegraphics[scale=0.4]{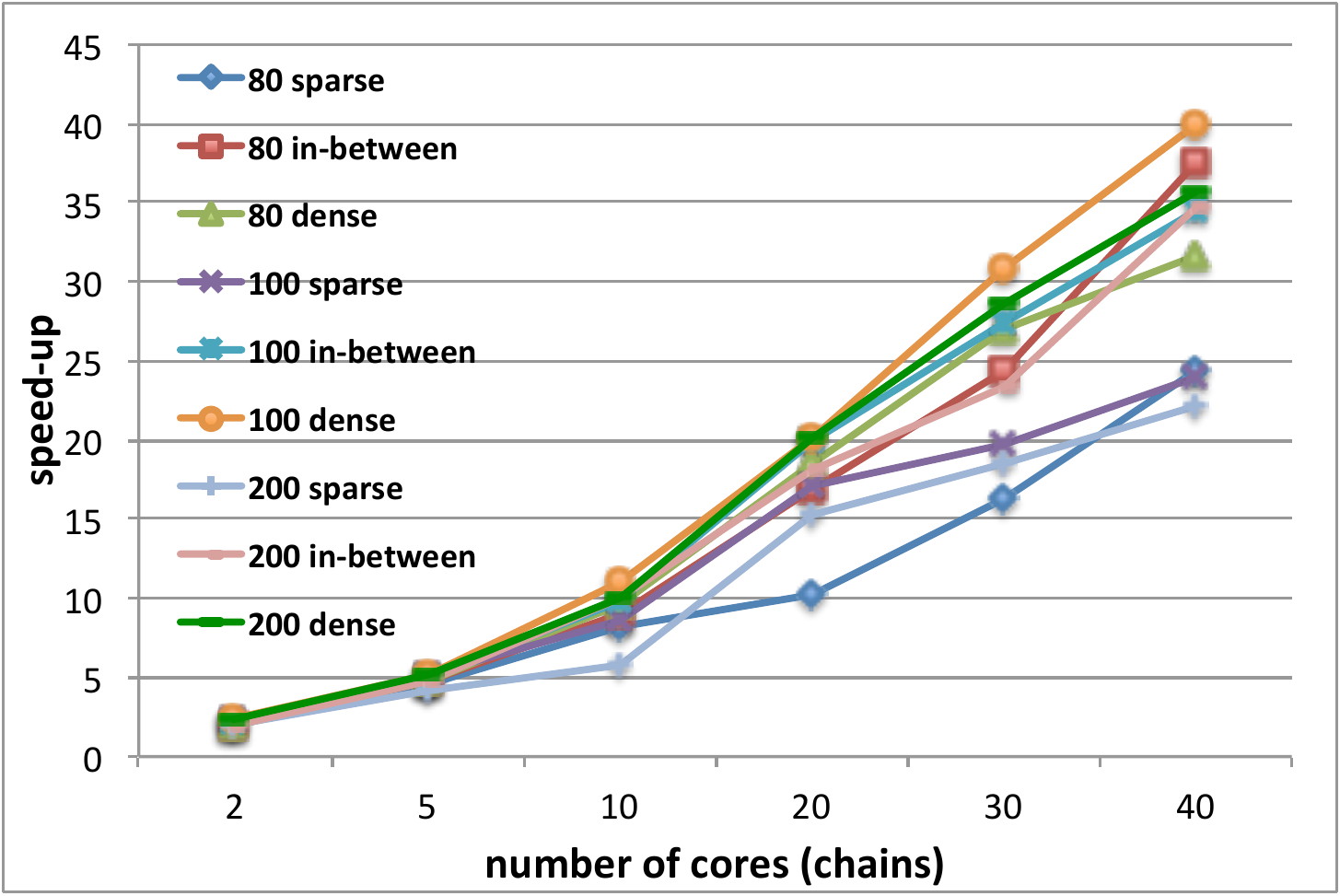}
    \caption{The two-state Markov chain approach (I): PBNs with 80, 100, 200 nodes. }
    \label{fig:ts-1}
  \end{subfigure}%
  \quad
  \begin{subfigure}[b]{0.48\textwidth}
    \centering
    \includegraphics[scale=0.4]{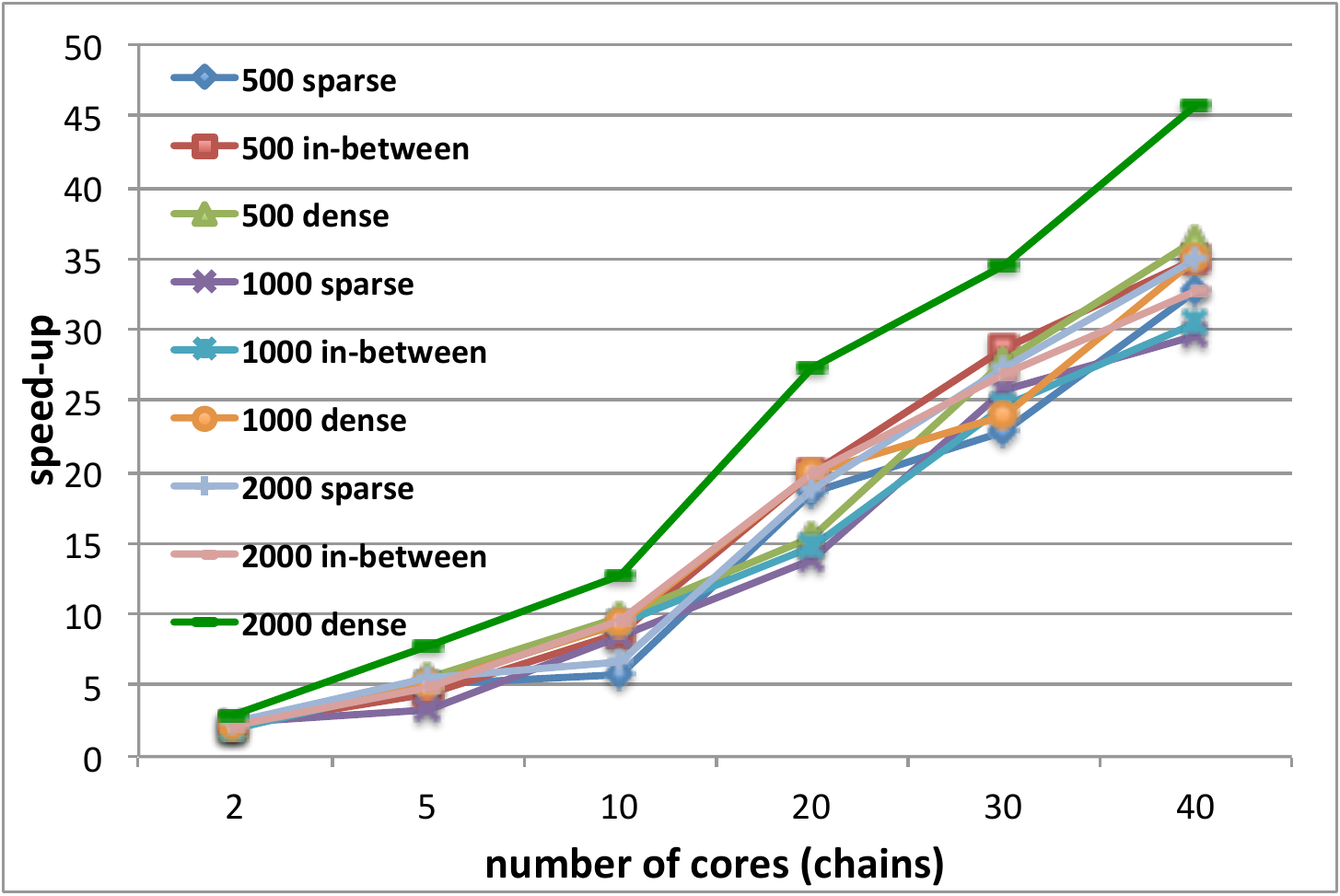}
    \caption{The two-state Markov chain approach (II): PBNs with 500, 1000, 2000 nodes. }
    \label{fig:ts-2}
  \end{subfigure}
  \begin{subfigure}[b]{0.48\textwidth}
    \centering
    \includegraphics[scale=0.4]{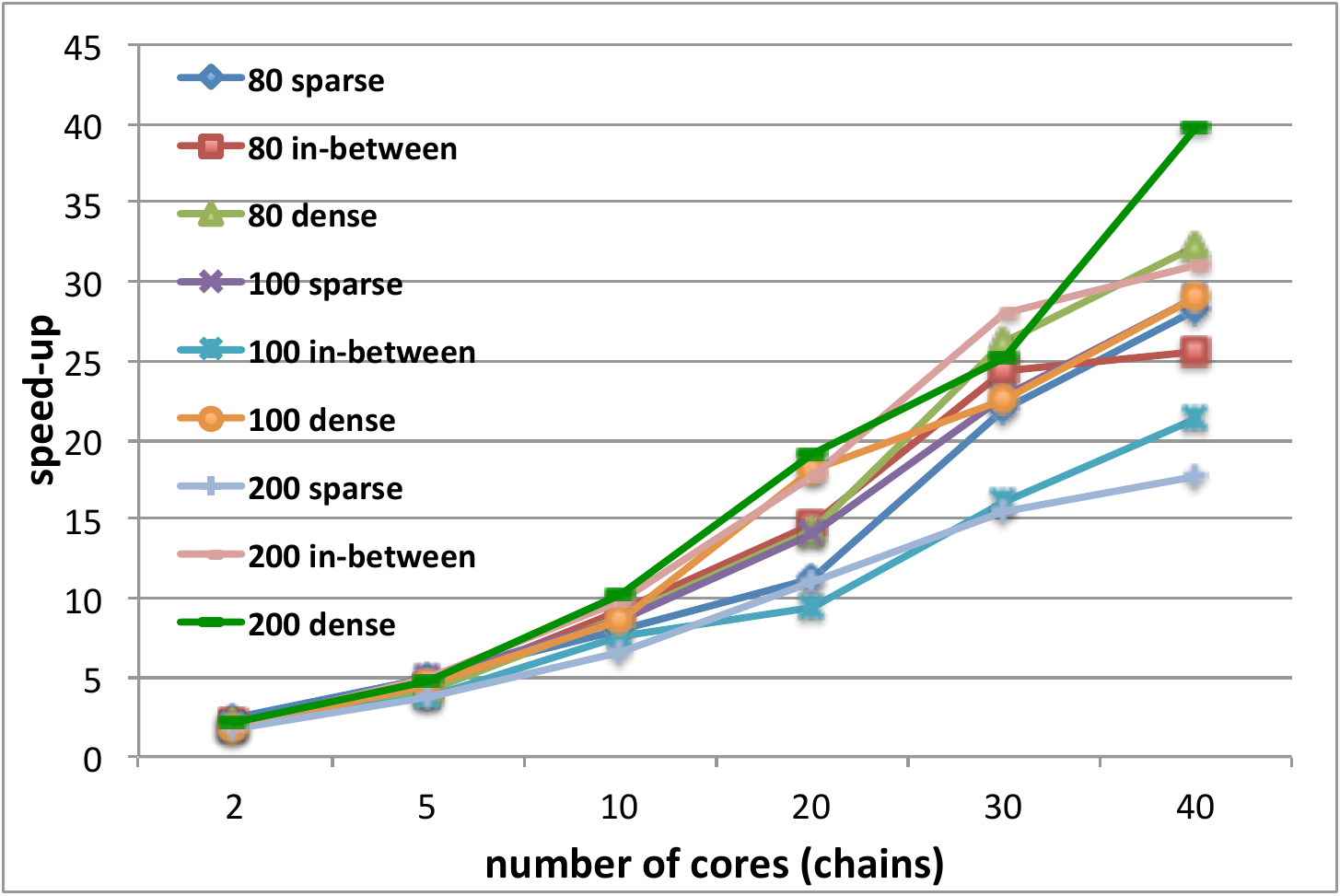}
    \caption{The Skart method (I): PBNs with 80, 100, 200 nodes. }
    \label{fig:sk-1}
  \end{subfigure}%
    \quad
  \begin{subfigure}[b]{0.48\textwidth}
    \centering
    \includegraphics[scale=0.4]{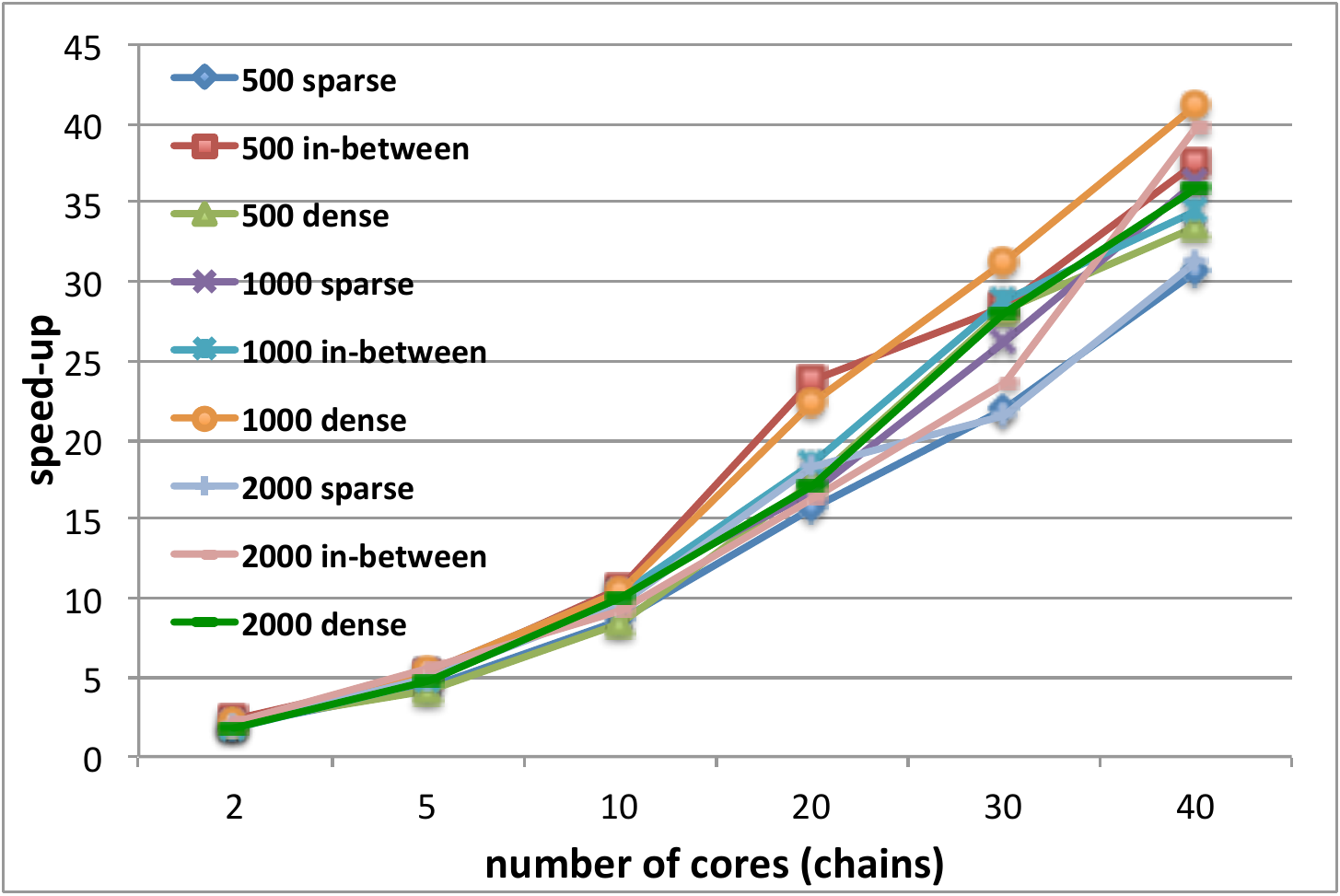}
    \caption{The Skart method (II): PBNs with 500, 1000, 2000 nodes.}
    \label{fig:sk-2}
  \end{subfigure}
  \caption{Speed-up of the parallelised algorithms.}
  \label{fig:performance}
\end{figure}

Figure~\ref{fig:ts-1} and Figure~\ref{fig:ts-2} present the speed-ups when analysing a single property of the 18 PBNs
with the parallelised two-state Markov chain approach.
Each speed-up is computed as $t_{sequential}/t_{parallel}$, where $t_{sequential}$ is the time cost
of the sequential two-state Markov chain approach and $t_{parallel}$ is the time cost of its parallelised version.
The parallelised approach is launched with different number of cores ranging in $\{2, 5, 10, 20, 30, 40\}$.
We see clearly from these figures that the speed-up is almost proportional to the number of cores.
Meanwhile, the speed-ups vary a lot in different models with 40 cores, e.g.,
we observe a speed-up about 46 in the case of the 2000-node dense model and a speed-up of 22 in the case of  the 200-node sparse model.

On the one hand, the required sample size varies in different runs due to the nature of the two-state Markov chain approach.
The speed-up can be bigger than the number of cores when the required sample size
in the parallelised run is smaller than what is required in the sequential run
--  this is actually the case where we obtain the speed-up of 46 for the 2000-node dense model.
On the contrary, the speed-up can be much smaller
than the number of cores when the required sample size in the parallelised run
is bigger than that required in the sequential run
-- this is the case where we obtain the speed-up of 31.7 for the 80-node dense model.
To show the affection of sample size,
we compute the speed-ups for the parallelised run with 40 cores after eliminating the affection of sample size
using the formula $sp*size_p/size_s$,
where $sp$ is the original speed-up computed with $t_{sequential}/t_{parallel}$,
$size_s$ is the sample size used for the sequential run,
and $size_p$ is the sample size used for the parallel run.
The results are shown in Table~\ref{tab:speed-up}(in the rows labelled with speed-up E).
On the other hand, the time cost also varies when checking a property for different models.
When the time cost of the sequential run is small,
the percentage of time spent in generating samples is also small.
As a consequence, the percentage of time the parallelised algorithm can reduce is small as well.
Therefore, the speed-up the parallelised algorithm can gain is small when the time cost of the sequential run is small
-- this is the case where we obtain the speed-up of 22 for the 200-node sparse model.
On the contrary, a larger speed-up is easy to obtain
when the time cost of the sequential run is big
-- this is the case where we obtain the speed-up of 46 for the 2000-node dense model.
We obtain similar speed-ups with the Skart method and
the speed-ups are presented in Figure~\ref{fig:sk-1} and Figure~\ref{fig:sk-2}.

Besides, we obtain maximum speed-ups with the use of 40 cores under the current hardware condition.
To illustrate this, 
we show in Table~\ref{tab:speed-up} more detailed information,
i.e., the time costs (in minutes),
the actual sample sizes (of millions), the speed-ups, and the speed-ups after eliminating the affection from the sample size.
Note that in order to make the results as accurate as possible,
all speed-ups are computed using the original time and size values we get from experiments,
not the approximate ones shown in Table~\ref{tab:speed-up}.
For the two-state Markov chain approach,
the speed-ups are greater than 30 for 14 out of 18 cases
and for the Skart method, the speed-ups are greater than 30 for 12 out of 18 cases.

Moreover,
we show in the next section that with the use of 40 cores, speed-ups between 19.67 and 33.04 are obtained for a 96-node PBN from a real biological system. 
\begin{table}[!t]
\center
\begin{tabular}{|c|c|r||r|r|r|r|r|r|r|r|r|}
\hline
 \multicolumn{3}{|c||}{node \# }&\multicolumn{3}{c|}{80}                                            & \multicolumn{3}{c|}{100}                                                 & \multicolumn{3}{c|}{200}                                                 \\ \hline
\multicolumn{3}{|c||}{density} &\multicolumn{1}{c|}{sparse} & \multicolumn{1}{c|}{in-between} & \multicolumn{1}{c|}{dense} & \multicolumn{1}{c|}{sparse} & \multicolumn{1}{c|}{in-between} & \multicolumn{1}{c|}{dense} & \multicolumn{1}{c|}{sparse} & \multicolumn{1}{c|}{in-between} & \multicolumn{1}{c|}{dense} \\ \hline
\multirow{5}{*}{\rotatebox[origin=c]{90}{two-state ~~~}} & \multirow{2}{*}{\parbox{0.6cm}{\centering time (m)}}&  seq.&    27.4	         &     50.7	               &       30.6	  &  30.4	   &           52.5	   & 	119.5		    &     18.0	 &  159.3	   &       171.0      \\ \cline{3-12}
&&par.&               1.1	&1.4	&1.0&	1.3	&1.5&	3.0&	0.8&	4.6&	4.8 \\ \cline{2-12}
&\multirow{2}{*}{\parbox{1.2cm}{\centering size (million)}} & seq.&  84.7	&105.3	&41.6&	70.2	&98.8&	117.7&	25.9&	134.6&	83.7 \\ \cline{3-12}
&&par.&    85.2	&105.0	&49.3&	70.3&	98.8	&117.5&	25.9&	134.7&	83.7 \\ \cline{2-12}
&\multicolumn{2}{c||}{speed-up} &  24.3&	37.5&	31.7&	23.9&	34.5&	40.0&	22.1&	34.7	&35.8             \\ \cline{2-12}
&\multicolumn{2}{c||}{speed-up E} &  24.5	&37.4	&37.7&	23.9	&34.5&	39.9&   22.2&	34.7&	35.8         \\ \hline
\multirow{4}{*}{\rotatebox[origin=c]{90}{Skart ~~~}}  & \multirow{2}{*}{\parbox{0.6cm}{\centering time (m)}}&  seq.&  29.9&	49.2&	31.7&	30.5	&47.6&	110.0&	16.7	&167.3	&176.9 \\ \cline{3-12}
&&par.&  1.1	&1.9	&1.0&	1.1&	2.2	&3.8&	0.9&	5.4&	4.4 \\ \cline{2-12}
&\multirow{2}{*}{\parbox{1.2cm}{\centering size (million)}} & seq.& 94.1&	110.2&	43.5	&72.7&	89.6	&110.1&	24.5	&141.2&	86.9 \\ \cline{3-12}
&&par.&  78.3	&109.7&	40.8&	69.6&	109.7	&123.5	&29.1&	131.1&	73.8  \\ \cline{2-12}
&\multicolumn{2}{c||}{speed-up} & 28.2&	25.5	&32.3&	28.9&	21.3&	28.9	&17.8&	31.0	&39.8 \\ \cline{2-12}
&\multicolumn{2}{c||}{speed-up E} &  23.4&	25.4	&30.3&	27.7&	26.1&	32.5&	21.1	&28.8&	33.8             \\ \hline\hline
 \multicolumn{3}{|c||}{node \# }&\multicolumn{3}{c|}{500}                                            & \multicolumn{3}{c|}{1000}                                                 & \multicolumn{3}{c|}{2000}                                                 \\ \hline
\multicolumn{3}{|c||}{density} &\multicolumn{1}{c|}{sparse} & \multicolumn{1}{c|}{in-between} & \multicolumn{1}{c|}{dense} & \multicolumn{1}{c|}{sparse} & \multicolumn{1}{c|}{in-between} & \multicolumn{1}{c|}{dense} & \multicolumn{1}{c|}{sparse} & \multicolumn{1}{c|}{in-between} & \multicolumn{1}{c|}{dense} \\ \hline
\multirow{4}{*}{\rotatebox[origin=c]{90}{two-state ~~~}} & \multirow{2}{*}{\parbox{0.6cm}{\centering time (m)}}&  seq.&  302.5&	556.9&	410.8&	211.3&	620.4&	1841.7&	111.0&	460.5&	643.1  \\ \cline{3-12}
&&par.&   9.3&	15.9&	11.4	&7.2	&20.3&	52.6&	3.2&	14.0&	14.0 \\ \cline{2-12}
&\multirow{2}{*}{\parbox{1.2cm}{\centering size (million)}} & seq.&   156.1	&198.3&	113.9&	75.8&	176.7&	297.6&	44.8	&114.1&	115.5 \\ \cline{3-12}
&&par.&   153.8	&204.8	&114.0	&75.6&	176.4&	302.1&	40.8	&113.5&	91.8   \\ \cline{2-12}
&\multicolumn{2}{c||}{speed-up} &  32.7	&34.9&	36.2	&29.5&	30.6&	35.0&	34.9&	32.8	&45.8 \\ \cline{2-12}
&\multicolumn{2}{c||}{speed-up E} &  32.2&	36.0&	36.2&	29.5&	30.5&	35.6&	31.8	&32.6&	36.4 \\\hline
\multirow{4}{*}{\rotatebox[origin=c]{90}{Skart ~~~~}}  & \multirow{2}{*}{\parbox{0.6cm}{\centering time (m)}}&  seq.& 278.5	&594.0&	394.2&	218.7&	671.3&	2095.5	&98.9&	467.5&	466.9 \\ \cline{3-12}
&&par.&  9.1	&15.9	&11.7&	6.0&	19.5&	51.0	&3.2&	11.7&	13.0 \\ \cline{2-12}
&\multirow{2}{*}{\parbox{1.2cm}{\centering size (million)}} & seq.& 144.1&	216.5&	114.9&	78.7&	185.8&	292.9&	40.1	&109.2&	94.1 \\ \cline{3-12}
&&par.& 134.9	&208.6&	117.4&	75.3&	184.4&	274.0&	38.4	&103.3	&86.4\\ \cline{2-12}
&\multicolumn{2}{c||}{speed-up} &30.6	&37.4	&33.6&	36.3&	34.5&	41.1&	31.3	&39.8&	35.8   \\ \cline{2-12}
&\multicolumn{2}{c||}{speed-up E} &  28.7&	36.0	&34.3	&34.7	&34.3&	38.5&	29.9&	37.7&	32.9 \\\hline
\end{tabular}
\caption{Speed-ups using 40 cores for Algorithm~\ref{alg:combined} and Algorithm~\ref{alg:skartparallel}.}
\label{tab:speed-up}
\end{table}

\subsection{Speed-up for checking multiple properties}
\label{ssec:speedupmultiproperties}
We have performed one influence analysis and two long-run sensitivity analyses of an apoptosis network
using the sequential two-state Markov chain approach in~\cite{MPY15}.
The apoptosis network contains 96 nodes;
one of the nodes, i.e., UV, can take on three values
and was refined as UV(1) and UV(2) in order to cast its original model
into the binary PBN framework.
The 96 nodes with 107 Boolean functions and their parameters, i.e., the selection probabilities of Boolean functions, were fitted to experimental data in~\cite{PAJ14}.
We took the 20 best fitted parameter sets and performed our analyses for them.
With an efficient implementation of a PBN simulator,
we managed to finish this analysis in an affordable amount of time.
Nevertheless, the analysis was still very expensive in terms of computation time
since the trajectories required were huge and we needed to check a number of properties.

In this work, we re-perform part of the influence analyses
by running the parallelised two-state Markov chain approach for checking 7 properties simultaneously with 40 cores.
In the influence analysis,
we aim to compute the \textit{long-term influences} on complex2 from each of its parent nodes: RIP-deubi, complex1, and FADD, in accordance with the definition in~\cite{SDKZ02}.
We consider both the case of UV(1) and UV(2) and hence we construct 2 PBNs for each of the 20 best fit parameter sets.
In total, we need to compute 7 different steady-state probabilities for 40 different PBNs.


Previously, we have applied the two-state Markov chain approach 280 times to finish the computation.
Using the parallelised version, we only need to perform the parallelised two-state Markov chain approach 40 times
since 7 properties for one PBN can be checked in one run.
In this evaluation, we perform the parallelised two-state Markov chain approach to
check the 7 properties of one of the 40 PBNs simultaneously and
show in Table~\ref{tab:ts-multi} the time cost (in minutes),
the actual sample size (of millions) we use and the speed-ups we obtain for checking
them with the sequential and parallelised algorithms.
To make the comparison complete,
we also perform the parallelised two-state Markov chain approach to check the 7 properties one by one
and show the results in Table~\ref{tab:ts-multi}.
The precision $r$, confidence level $s$, and steady-state convergence parameter $\epsilon$ in this
experiment are set to $10^{-5}$, 0.95 and $10^{-10}$, respectively.
The speed-ups for checking a single property are computed similarly as in Figure~\ref{fig:performance};
while the speed-up for checking the 7 properties is computed with $\sum_{i=1}^{7}{t_i}/t_{multi}$,
where $t_i$ is the time cost for checking the $i$-th property with the sequential algorithm
and $t_{multi}$ is the time cost for checking the 7 properties simultaneously with the parallelised algorithm.
From Table~\ref{tab:ts-multi},
the parallelised algorithm obtains a speed-up between 19.67 and 33.04 for checking a single property
and a speed-up of 52.88 for checking 7 properties.
By reuse of generated samples, the sample size is also reduced by 3.45 times from 1563.05 to 452.74 million.

\begin{table}[!t]
\centering
\begin{tabular}{|c|r|r||c|r|r||r|}
\hline
\multicolumn{3}{|c||}{sequential}                                                                                                   & \multicolumn{3}{c|}{parallelised}                                                                                                 & \multicolumn{1}{c|}{\multirow{3}{*}{speed-up}} \\ \cline{1-6}
property \# & \multicolumn{1}{c|}{\begin{tabular}[c]{@{}c@{}}sample size\\ (million)\end{tabular}} & \multicolumn{1}{c||}{time (m)} & property \# & \multicolumn{1}{c|}{\begin{tabular}[c]{@{}c@{}}sample size\\ (million)\end{tabular}} & \multicolumn{1}{c|}{time(m)} & \multicolumn{1}{c|}{}                          \\ \hline
1           & 146.99                                                                               & 53.30                         & 1           & 147.20                                                                               & 2.31                         & 23.04                                          \\ \hline
2           & 454.14                                                                               & 174.25                        & 2           & 461.58                                                                               & 6.81                         & 25.58                                          \\ \hline
3           & 253.45                                                                               & 97.77                         & 3           & 253.60                                                                               & 3.86                         & 25.32                                          \\ \hline
4           & 48.81                                                                                & 16.71                         & 4           & 50.11                                                                                & 0.73                         & 22.83                                          \\ \hline
5           & 305.35                                                                               & 120.33                        & 5           & 335.85                                                                               & 5.38                         & 22.39                                          \\ \hline
6           & 50.21                                                                                & 17.65                         & 6           & 51.31                                                                                & 0.90                         & 19.67                                          \\ \hline
7           & 255.17                                                                               & 99.75                         & 7           & 263.39                                                                               & 3.02                         & 33.04                                          \\ \hline\hline
total       & 1563.05                                                                              & 579.75                         & 1-7         & 452.74                                                                               & 10.96                        & 52.88                                          \\ \hline
\end{tabular}
\caption{Performance comparison on multiple properties
with the two-state Markov chain approach.}
\label{tab:ts-multi}
\end{table}

\section{Conclusion and Future Work}
\label{sec:future}
In this paper, we proposed to combine the German \& Rubin method with two
statistical methods, i.e.,
the two-state Markov chain approach and the Skart method,
to reduce the time cost for computing steady-state probabilities of large PBNs.
We showed with experiments that the proposed combinations can reduce the time cost of the original sequential methods significantly.


Our parallelised algorithms work well on multiple-core CPU architecture. However, the scalability
of multiple-core CPU based parallelisation is often restricted by the CPU architecture since the
number of processing units in a~CPU is usually small. On the contrary, GPUs often contain
thousands of processing units and GPUs achieve high performance when thousands of threads execute
concurrently~\cite{MP09}. Parallelising those algorithms with GPU based architecture will
potentially lead to further speed-ups.

\bibliographystyle{splncs}
\bibliography{pbn}

\end{document}